\documentstyle[12pt]{article}
\newlength{\extraspace}
\newlength{\extraspaces}
\newcommand{\be}{\begin{equation}
\addtolength{\abovedisplayskip}{\extraspaces}
\addtolength{\belowdisplayskip}{\extraspaces}
\addtolength{\abovedisplayshortskip}{\extraspace}
\addtolength{\belowdisplayshortskip}{\extraspace}}
\newcommand{\ee}{\end{equation}}
\newcommand{\ba}{\begin{eqnarray}
\addtolength{\abovedisplayskip}{\extraspaces}
\addtolength{\belowdisplayskip}{\extraspaces}
\addtolength{\abovedisplayshortskip}{\extraspace}
\addtolength{\belowdisplayshortskip}{\extraspace}}
\newcommand{\ea}{\end{eqnarray}}
\newcommand{\nonu}{\nonumber \\[2mm]}
\newcommand{\is}{& \!\! = \!\! &}
\newcommand{\twomatrixd}[4]{{\left(\begin{array}{cc}
\displaystyle #1 & \displaystyle #2\\[2mm]
\displaystyle  #3  & \displaystyle #4 \end{array}\right)}}
\newcommand{\ie}{{\it i.e.\ }}

\newcommand{\half}{{\textstyle{1\over 2}}}

\newcommand{\Z}{{\bf Z}}           
\newcommand{\R}{{\bf R}}           
  
\newcommand{\cH}{{\cal H }}            
\newcommand{\cF}{{\cal F }}            
\newcommand{\cZ}{{\cal Z }}            
            
\newcommand{\ra}{\rightarrow}

\newcommand{\qbar}{{\overline{q}}}

\newcommand{\Lbar}{{\overline{L}}}

\newcommand{\ext}{{\raisebox{.2ex}{$\textstyle \bigwedge$}}}

\input epsf
\newcounter{fignum}
\newcommand{\figuurnum}{\arabic{fignum}}
\newcommand{\figuur}[3]{
\addtocounter{fignum}{1}
\addcontentsline{lof}{figure}{\protect
\numberline{\arabic{section}.\arabic{fignum}}{#3}}
\hspace{-3mm}{\it fig.}\ \figuurnum.
\begin{figure}[t]\begin{center}
\leavevmode\hbox{\epsfxsize=#2 \epsffile{#1.eps}}\\[3mm]
\parbox{11cm}{\small \bf Fig.\ \figuurnum: \it #3}
\end{center} \end{figure}\hspace{-1.5mm}}

\newcommand{\newsection}[1]{
\vspace{15mm}
\pagebreak[3]
\addtocounter{section}{1}
\setcounter{equation}{0}
\setcounter{subsection}{0}
\setcounter{footnote}{0}
\setcounter{fignum}{0}
\begin{center}
{\sc \thesection. #1}
\end{center}
\nopagebreak
\medskip
\nopagebreak}
\newcommand{\newsubsection}[1]{
\vspace{1cm}
\pagebreak[3]
\addtocounter{subsection}{1}
\addcontentsline{toc}{subsection}{\protect
\numberline{\arabic{section}.\arabic{subsection}}{#1}}
\noindent{ \it \thesubsection. #1}                 
\nopagebreak
\vspace{2mm}
\nopagebreak}
\newcommand{\Tr}{{\rm Tr}}
\newcommand{\MM}{{M}}
\newcommand{\SNM}{S^N\!\MM}
\newcommand{\Sym}{S}
\newcommand{\Dslash}{{D \hspace{-8pt} \slash}}
\newcommand{\cM}{{\cal M}}
\newcommand{\cI}{{\cal I}}

\begin{document}
\addtolength{\baselineskip}{.5mm}
\begin{flushright}
August 1996\\
{\sc hep-th}/9608096\\
{\sc cern-th}/96-222\\
{\sc itfa}/96-31\\
{\sc yctp-p}16-96
\end{flushright}
%\vspace{-.3cm}
\thispagestyle{empty}
\begin{center}
{\large\sc{Elliptic Genera of Symmetric Products\\[3mm]
and Second Quantized Strings}}\\[15mm]

{\sc Robbert Dijkgraaf}${}^1$, {\sc Gregory Moore}${}^2$,\\[2mm]
{\sc Erik Verlinde}${}^3$ and {\sc Herman Verlinde}${}^4$\\[9mm]
${}^1${\it Mathematics Dept, Univ. of Amsterdam, 1018 TV Amsterdam}\\[4mm]
${}^2${\it Physics Dept, Yale University, New Haven, CT 06520}\\[4mm]
${}^3${\it Theory Division, CERN, CH-1211 Geneva 23, and}\\[.1mm]
{\it Inst. for Theor. Physics, University of Utrecht, 3508 TA Utrecht}\\[4mm]
${}^4${\it Inst. for Theor. Physics, Univ. of Amsterdam, 1018 XE Amsterdam} 
\\[24mm]

Abstract
\end{center}
In this note we prove an identity that equates the elliptic genus
partition function of a supersymmetric sigma model on the $N$-fold
symmetric product $M^N/S_N$ of a manifold $M$ to the partition
function of a second quantized string theory on the space $M \times
S^1$. The generating function of these elliptic genera is shown to be
(almost) an automorphic form for $O(3,2,\Z)$. In the context of
D-brane dynamics, this result gives a precise computation of the free
energy of a gas of D-strings inside a higher-dimensional brane.

\newpage

\newsection{The Identity}

Let $\MM$ be a K\"ahler manifold. In this note we will consider
the partition function of the supersymmetric sigma model defined
on the $N$-fold symmetric product $\SNM$ of $\MM$, which is the orbifold
space
\be
\SNM = \MM^N/S_N
\ee
with $S_N$ the symmetric group of $N$ elements.  The genus one
partition function depends on the boundary conditions imposed on the
fermionic fields. For definiteness, we will choose the boundary
conditions such that the partition function $\chi(\SNM;q,y)$ coincides
with the elliptic genus \cite{book,ell}, which is defined as the trace
over the Ramond-Ramond sector of the sigma model of the evolution
operator $q^{H}$ times $(-1)^{F} y^{F_L}$. Here $q$ and $y$ are
complex numbers and $F=F_L+ F_R$ is the sum of the left- and
right-moving fermion number.  (See the Appendix for background.)  In
particular,
\be
\chi(\MM ; q,y) = \Tr_{\strut \! \cH(\MM)} 
(-1)^F y^{F_L} q^H
\ee
with $H = {L_0 - {c\over 24}}$.  Of the right-moving sector only the
R-ground states contribute to the trace.

We will prove here an identity, conjectured in \cite{dyon}, that
expresses the orbifold elliptic genera of the symmetric product
manifolds in terms of that of $\MM$ as follows
\footnote{In case we have more than one conserved
quantum number such as $F_L$, the index $\ell$ becomes a multi-index 
and the denominator
on the RHS of (\ref{identity}) becomes a general product formula as
appears in the work of Borcherds \cite{borcherds}, see also
\cite{harveymoore}.}
\be
\label{identity}
\sum_{N=0}^\infty p^N \chi(\SNM;q,y) = \prod_{n>0,m\geq 0,\ell}
{1\over (1-p^n q^m y^\ell)}{}_{c(nm,\ell)}
\ee 
where the coefficients $c(m,\ell)$ on the right-hand side are defined via
the expansion 
\be
\label{single}
\chi(\MM; q,y) = 
\sum_{m\geq 0,\ell} c(m,\ell) q^m y^\ell.
\ee
The proof of this identity follows quite directly from borrowing
standard results about orbifold conformal field theory
\cite{orbifold}, and generalizes the orbifold Euler number computation
of \cite{hirzebruch} (see also \cite{vafawitten}).  Before presenting
the proof, however, we will comment on the physical interpretation of
this identity in terms of second quantized string theory.

\newsubsection{String Theory Interpretation}

Each term on the left-hand side with given $N$ can be thought of as
the left-moving partition sum of a single (non-critical)
supersymmetric string with %target-space $\SNM \times S^1$.  target
space-time $\SNM \times S^1\times \R$.  This string is wound once
around the $S^1$ direction, and in the light-cone gauge its
transversal fluctuations are described by the supersymmetric
sigma-model on $\SNM$. The right-hand side, on the other hand, can be
recognized as a partition function of a large Fock space, made up from
bosonic and fermionic (depending on whether $c(nm,\ell)$ is positive
or negative) creation operators $\alpha^I_{n,m,\ell}$ with
$I=1,2,\ldots,|c(nm,\ell)|$. This Fock space is identical to the one
obtained by second quantization of the left-moving sector of the
string theory on the space $\MM \times S^1$.  In this correspondence,
the oscillators $\alpha^I_{n,m,\ell}$ create string states with
winding number $n$ and momentum $m$ around the $S^1$.  The number of
such states is easily read off from the single string partition
function (\ref{single}). In the light-cone gauge we have the level
matching condition
\be
\label{level}
L_0 - \overline{L}_0 = m n,
\ee
and since $\overline{L}_0 = 0$, this condition implies that the left-moving 
conformal dimension is equal to $h = mn$. Therefore, according to 
(\ref{single}) the number of
single string states with winding $n$, momentum $m$ and $F_L = \ell$ is 
indeed given by $|c(nm,\ell)|$. (Strictly speaking, the elliptic genus
counts the number of bosonic minus fermionic states at each oscillator
level. Because of the anti-periodic boundary condition in the
time direction for the fermions, only the net number contributes 
in the space-time partition function (\ref{identity}). )

The central idea behind the proof of the above identity is that the
partition function of a single string on the symmetric product $\SNM$
decomposes into several distinct topological sectors, corresponding to
the various ways in which a once wound string on $\SNM \times S^1$ can
be disentangled into separate strings that wind one or more times
around $\MM \times S^1$. To visualize this correspondence, it is
useful to think of the string on $\SNM \times S^1$ as a map that
associates to each point on the $S^1$ a collection of $N$ points in
$\MM$. By following the path of these $N$ points as we go around the
$S^1$, we obtain a collection of strings on $\MM \times S^1$ with
total winding number $N$, that reconnect the $N$ points with
themselves.  Since all permutations of the $N$ points on $\MM$
correspond to the same point in the symmetric product space, the
strings can reconnect in different ways labeled by conjugacy classes
$[g]$ of the permutation group $S_N$. The factorization of $[g]$ into
a product of irreducible cyclic permutations $(n)$ determines the
decomposition into several strings of winding number $n$. (See
\figuur{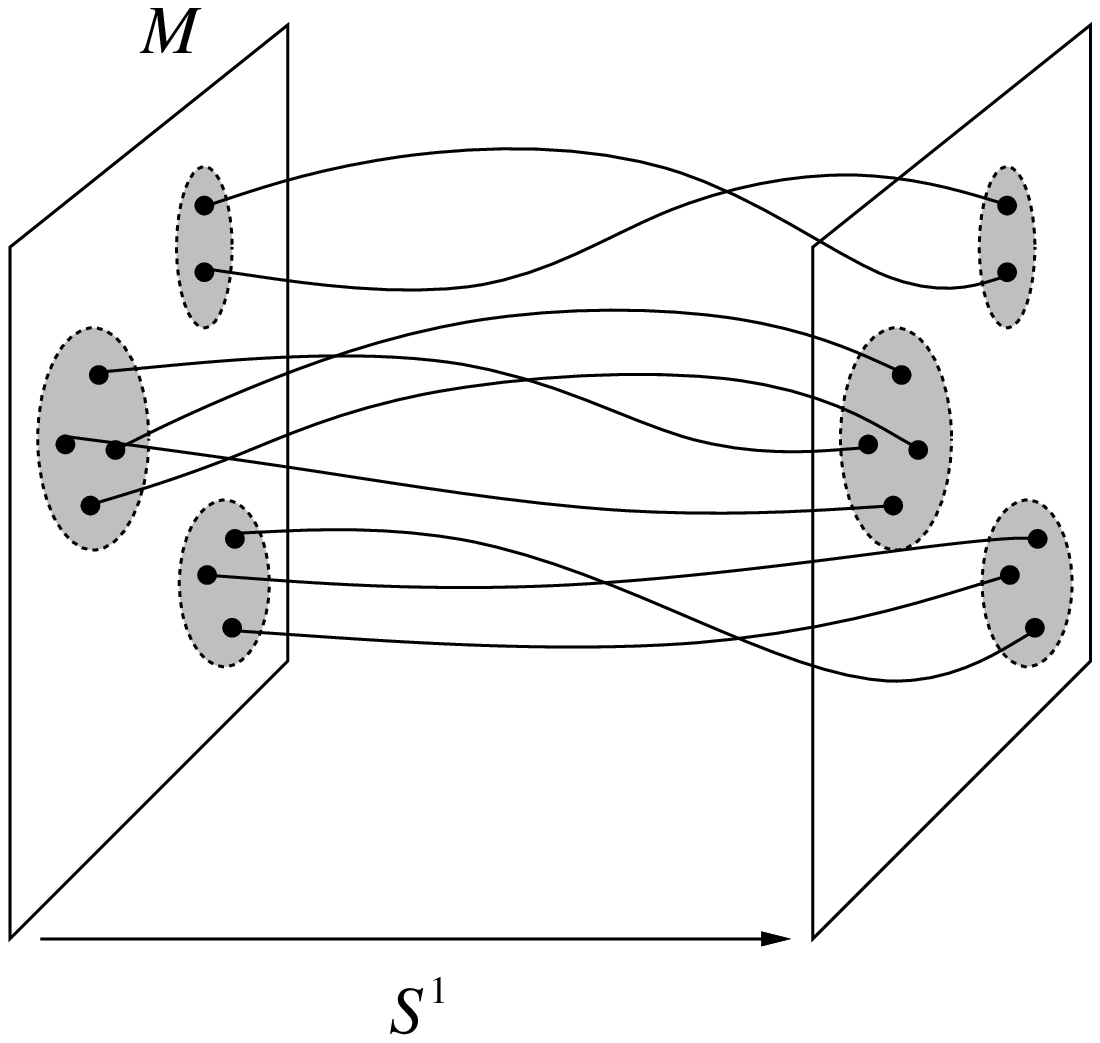}{8.0cm}{ The string configuration corresponding to a
twisted sector by a given permutation $g \in S_N$. The string
disentangles into seperate strings according to the factorization of
$g$ into cyclic permutations.})  The combinatorical description of the
conjugacy classes, as well as the appropriate symmetrization of the
wavefunctions, are both naturally accounted for in terms of a second
quantized string theory.

\newcommand{\ltimes}{\times \hspace{-6pt} 
\raisebox{.2ex}{\mbox{$\scriptscriptstyle |$}} \;}

\newsection{The Proof}

The Hilbert space of an orbifold field theory \cite{orbifold} 
is decomposed into 
twisted sectors $\cH_g$, that are labelled by the conjugacy classes
$[g]$ of the orbifold group, in our case the symmetric group $S_N$. 
Within each twisted sector, one only keeps the 
states invariant under the centralizer subgroup $C_g$ of $g$. 
We will denote this $C_g$ invariant subspace by $\cH_g^{C_g}$. 
Thus the total orbifold Hilbert space takes the form
\be
\cH(\SNM) = \bigoplus_{[g]} \cH_g^{C_g}.
\ee
For the symmetric group, the conjugacy classes $[g]$ are characterized
by partitions $\{N_n\}$ of $N$ 
\be
\sum_{n} n N_n = N,
\ee
where $N_n$ denotes the multiplicity of the cyclic permutation $(n)$
of $n$ elements in the decomposition of $g$
\be
[g] = (1)^{N_1}(2)^{N_2} \ldots (s)^{N_s}.
\ee
The centralizer subgroup
of a permutation $g$ in this conjugacy class takes the form
\be
C_g = S_{N_1} \times (S_{N_2} \ltimes \Z^{N_2}_2) \times \ldots 
(S_{N_s} \ltimes \Z^{N_s}_s). 
\ee
Here each subfactor $S_{N_n}$ permutes the $N_n$ cycles $(n)$, while each
subfactor $\Z_n$ acts within one particular cycle $(n)$.

Corresponding to the above decomposition of $[g]$ into irreducible
cyclic permutations, we can decompose each twisted sector
$\cH_g^{C_g}$ into the product over the subfactors $(n)$ of $N_n$-fold
symmetric tensor products of appropriate smaller Hilbert spaces
$\cH^{\Z_n}_{(n)}$
\be
\label{hdecomposition}
{\cH}^{C_g}_g = \bigotimes_{n>0} \, \Sym^{N_n} \cH_{(n)}^{\Z_n},
\ee
where we used the following notation for (graded) symmetric tensor products
\be
\Sym^N\cH = \left(\underbrace{\cH \otimes \ldots \otimes \cH}_{N\ times}
\right)^{S_N}.
\ee
Here the symmetrization is assumed to be compatible with the grading
of $\cH$. In particular for pure odd states $S^N$ 
corresponds to the exterior product $\ext^N$.

The Hilbert spaces $\cH_{(n)}^{\Z_n}$ in (\ref{hdecomposition}) denote
the $\Z_n$ invariant subsector of the Hilbert space $\cH_{(n)}$ of a
single string on $M\times S^1$ with winding number $n$. We can
represent $\cH_{(n)}$ as the Hilbert space of the sigma model of $n$
coordinate fields $X_i(\sigma) \in \MM$ with the cyclic boundary
condition
\be
\label{boundaryc}
\qquad X_i(\sigma+2\pi) = X_{i+1}(\sigma), \qquad i \in (1, \ldots, n). 
\ee
The group $\Z_n$, acting on the Hilbert  space $\cH_{(n)}$, is 
generated by the 
cyclic permutation
\be
\label{omega}
\omega : \ X_i \rightarrow X_{i+1}.
\ee
We can glue the $n$ coordinate fields $X_i(\sigma)$ together into one
single field $X(\sigma)$ defined on the interval $0\leq
\sigma \leq 2\pi n$. Hence, relative to the string with winding number 
one, the oscillators of the long string that generate $\cH_{(n)}$ have a 
fractional  ${1\over n}$ moding. The $\Z_n$-invariant subspace $\cH_{(n)}^{\Z_n}$
consists of those states in $\cH_{(n)}$ for which the fractional oscillator 
numbers combined add up to an integer. We will make use of this observation
in the next subsection.

\newsubsection{Partition Function of a Single String}

The  elliptic genus of $\SNM$ can now be computed by taking the
trace over the Hilbert space in the various twisted sectors.
We introduce the following notation:
\be
\chi(\cH;q,y) = 
\Tr_{\strut \cH} (-1)^F y^{F_L} q^H 
\ee
for every (sub)Hilbert space $\cH$ of a supersymmetric sigma-model.
Note that
\ba
\chi(\cH \oplus \cH';q,y) \is \chi(\cH;q,y) + \chi(\cH';q,y),\nonu
\chi(\cH \otimes \cH';q,y) \is \chi(\cH;q,y) \cdot \chi(\cH';q,y).
\label{blurp}
\ea
These identities will be used repeatedly in the following.

As the first step we will now compute the elliptic genus of the
twisted sector $\cH_{(n)}$. This is the left-moving partition sum of a
single string with winding $n$ on $\MM \times S^1$. As we have
explained, its elliptic genus can be simply related to that of a
string with winding number one via a rescaling $q \rightarrow
q^{1\over n}$
\be
\chi(\cH_{(n)};q,y) = \chi(\cH; q^{1\over n},y) = 
\sum_{m\geq 0,\ell} c(m,\ell) q^{m\over n} 
y^\ell
\ee
This rescaling accounts for the fractional $1\over n$ moding of the 
string oscillation numbers. 

The projection on the $\Z_n$ invariant sector is implemented by
insertion of the projection operator $P = {1\over n} \sum_k \omega^k$,
with $\omega$ as defined in (\ref{omega}).
\be
\chi(\cH_{(n)}^{\Z_n}; q,y) = {1\over n} \sum_{k=0}^{n-1}
\Tr_{\strut \cH_{(n)}} 
\omega^k
(-1)^F y^{F_L} q^H .
\ee
Since the boundary condition (\ref{boundaryc}) on the Hilbert space
$\cH_{(n)}$ represents a $\Z_n$-twist by $\omega$ along the $\sigma$
direction, the operator insertion of $\omega$ in the genus one
partition sum can in fact be absorbed by performing a modular
transformation $\tau \rightarrow \tau+1$, which amounts to a
redefinition $q^{1\over n} \rightarrow q^{1\over n} e^{2\pi i\over
n}$.\footnote{The redefinition $\tau \rightarrow \tau+1$ means that
the periodic boundary condition in the time direction is composed with
a space-like translation $\sigma \rightarrow \sigma + 2\pi$. According
to (\ref{boundaryc}) and (\ref{omega}) this indeed results in an extra
insertion of the operator $\omega$ into the trace.}  Thus we can write
\ba
\chi(\cH_{(n)}^{\Z_n}; q,y) \is {1\over n} \sum_{k=0}^{n-1}
\chi(\cH ; q^{1\over n}e^{{2\pi i k}\over n}, y) \nonu
\is \sum_{m\geq 0,\ell} c(mn,\ell) q^{m} y^\ell.
\ea

\newsubsection{Symmetrized products}

The next step is to consider the partition function for the
symmetrized tensor products of the Hilbert spaces
$\cH^{Z_n}_{(n)}$. We need the following result: 
If $\chi(\cH;q,y)$ has the expansion
\be
\chi(\cH;q,y)=\sum_{m,\ell} d(m,\ell) q^m y^\ell,
\ee
then we want to show that the partition function of the symmetrized 
tensor products of $\cH$ is given by the generating function
\be
\label{A}
\sum_{N\geq 0} p^N \chi(\Sym^N\cH; q,y) =
\prod_{m,\ell} {1\over (1-p q^m y^\ell)}{}_{d(m,\ell)}
\ee 
This identity is most easily understood in terms of second
quantization.  The sum over symmetrized products of $\cH$ is described
by a Fock space with a generator for every state in $\cH$, where
states with negative ``multiplicities'' $d(m,\ell)$ are identified as
fermions.  The usual evaluation of the partition function in a Fock
space then results in the RHS of equation (\ref{A}).

In more detail, we can interpret the elliptic genus as computing the
(super)dimension\footnote{We define $\dim V= \Tr_V (-1)^F=d^+-d^-$,
where $d^\pm$ are the dimensions of the even and odd subspaces $V^\pm$
in the decomposition $V=V^+ \oplus V^-$.} of vector spaces
$V_{m,\ell}$
\be
d(m,\ell) = \dim V_{m,\ell}.
\ee
We then evaluate
\ba
\sum_{N\geq 0} p^N \chi( \Sym^N\cH ; q,y)\is
\sum_{N\geq 0} \, p^N \!
\sum_{m_1\,\ldots,m_{{}_{\! N}}\atop \ell_1,\ldots,
\ell_{{}_{\! N}}}
\dim\left(V_{m_1,\ell_1} \otimes \cdots \otimes 
V_{m_{{}_{\! N}},\ell_{{}_{\! N}}}\right)^{S_N} 
q^{m_1+\ldots + m_{{}_{\! N}}} y^{\ell_1+ \ldots + 
\ell_{{}_{\! N}}}\nonu
\is \sum_{N\geq 0}\, p^N \! \sum_{N_{m,\ell}\atop \sum N_{m,\ell}=N}
\prod_{m,\ell} \left(q^m y^\ell\right)^{N_{m,\ell}}
\dim \left(\Sym^{N_{m,\ell}} V_{m,\ell}\right) \nonu
\is \prod_{m,\ell}\; \sum_{N\geq 0}\, p^N \left(q^m y^\ell\right)^N
\dim \left(\Sym^N V_{m,\ell}\right).
\ea
Using the identity
\be
\dim \left(\Sym^N V_{m,\ell}\right) = { d(m,\ell) + N - 1 \choose N},
\ee
where the RHS is defined as $(-1)^N {|d(m,\ell)|\choose N}$ for negative
$d(m,\ell)$, gives the desired result.

\newsubsection{Combining the ingredients}

The proof of our main identity follows from combining the results of
the previous two subsections. Our starting point has been the fact
that the Hilbert space of the orbifold field theory has a
decomposition in terms of twisted sectors as
\be
\cH(\SNM) = \bigoplus_{\sum n N_n = N} \bigotimes_{n>0} \; \Sym^{N_n}\! 
\cH_{(n)}^{\Z_n}.
\ee
Physically speaking, the right-hand side describes the Hilbert space
of a second quantized string theory with $N_n$ the number of strings
with winding number $n$.

With this form of the Hilbert space $\cH(\SNM)$, we find for the partition 
function
\ba
\label{junk}
\sum_{N\geq 0} p^N \chi(\SNM;q,y) \is 
\sum_{N\geq 0} \, p^N \! \sum_{N_n \atop \sum n N_n = N} \; \prod_{n>0} 
\chi( \Sym^{N_n} \cH_{(n)}^{\Z_n}; q,y) \nonu
\is \prod_{n>0} \sum_{N\geq 0} p^{nN}  \chi(\Sym^N \cH_{(n)}^{\Z_n}; q,y).
\ea
Here we used repeatedly the identities (\ref{blurp}).  In order to
evaluate the elliptic genera of the symmetric products, we apply the
result (\ref{A}) of the previous subsection to the Hilbert space
$\cH_{(n)}^{\Z_n}$, which gives
\be
\sum_{N\geq 0} p^N \chi( \Sym^N\cH_{(n)}^{\Z_n} ; q,y) =
\, \prod_{m\geq 0,\ell} {1\over (1-p q^m y^\ell)}{}_{c(mn,\ell)}
\label{result}
\ee 
If we insert this into (\ref{junk}) we get our final identity
\be
\label{final}
\sum_{N\geq 0} p^N \chi(\SNM;q,y) =
\prod_{n>0,m\geq 0,\ell} {1\over (1-p^n q^m y^\ell)}{}_{c(mn,\ell)}
\ee
which concludes the proof.
 
\newsection{One-Loop Free Energy}

In this section we will discuss some properties of our identity. 
For convenience we will assume here that the space $M$ is a Calabi-Yau
manifold, so that the sigma-model defines a $N=2$ superconformal field 
theory. For the elliptic genus this implies that it transforms as
a modular form. 
 
We have argued that the quantity on the right-hand side of (\ref{final})
\be
\label{zee}
\cZ(p,q,y) = 
\prod_{n>0,m,\ell} {1\over (1-p^n q^m y^\ell)}{}_{c(mn,\ell)}
\ee
has an interpretation as the partition function of a second quantized
string theory with target space $M \times S^1$. This identification
was based on the fact that $\cZ$ has the form of the trace over a free
field Fock space generators by oscillators $\alpha^I_{n,m,l}$ with $I
=1,\ldots,|c(nm,\ell)|$, i.e. one oscillator for each first quantized
string state. We will now comment on the path integral derivation of
this expression.

Since we are dealing with a free string 
theory, we should be able to take the logarithm of the partition sum
\be
\label{free}
\cF(p,q,y) = \log \cZ(p,q,y)
\ee
and obtain an interpretation of $\cF$ as the one-loop free energy of a
single string. From a path-integral perspective, this free energy is
obtained by summing over irreducible one-loop string amplitudes.  The
time coordinate of the target space is taken to be compactified (since
the partition function is defined as a trace) and thus the irreducible
one loop string amplitudes are described in terms of all possible maps
of $T^2$ into the Euclidean target space-time $M\times T^2$. From this
point of view the parameters $p,q,y$ obtain the interpretation as
moduli of the target space two-torus.  We can introduce parameters
$\rho, \sigma,\upsilon$ via
\be
p=e^{2\pi i\rho},\quad q=e^{2\pi i\sigma},\quad y=e^{2\pi i \upsilon}.
\label{exp}
\ee
Here $\rho$ and $\sigma$ determine the complexified K\"ahler form and
complex structure modulus of $T^2$ respectively, whereas $\upsilon$ 
parametrizes the $U(1)$ bundle on $T^2$ corresponding to $F_L$. 

\newsubsection{Instanton sums and Hecke operators}

We will now show that the logarithm $\cF$ of the partition function 
(\ref{zee}) indeed has the
interpretation of a one-loop free energy for a string on $M\times T^2$. 
First we compute
\ba
\cF(p,q,y) \is - \sum_{n>0,m,\ell} c(nm,\ell)
 \log\left( 1 - p^n q^m y^\ell \right) \nonu
\is \sum_{n>0,m,\ell, k>0}{1\over k}  c(nm,\ell) 
p^{kn} q^{km} y^{k\ell}
\nonu
\is \sum_{N>0} p^N \sum_{kn=N} {1\over k} \sum_{m,\ell}  
c(nm,\ell) q^{km} y^{k\ell}.
\label{F}
\ea
To write this expression in a more convenient form, it is useful to
recall the definition of the Hecke operators $T_N$. (For more details
on Hecke operators see e.g.\ \cite{hecke}.) In general, the Hecke
operator $T_N$ acting on a weak Jacobi form\footnote{See the Appendix
for the definition of a Jacobi form.} $\phi(\tau,z)$ of weight zero
and index $r$ produces a weak Jacobi form $T_N\phi$ of weight zero and
index $Nr$, defined as follows
\be
\label{defitn}
T_N\phi(\tau,z) = \sum_{ad=N \atop b\ {\rm mod}\ d} {1\over N}
\,\phi\left({a\tau + b\over d},az\right).
\ee
Hence if $\phi(\tau,z)$ has a  Fourier expansion
\be
\phi(\tau,z) = \sum_{m\geq 0,\ell} c(m,\ell) q^m y^\ell,
\ee
then $T_N\phi(\tau,z)$ takes the form
\be
\label{deftn}
T_N\phi(\tau,z) = \sum_{ad=N} {1\over a} \sum_{m\geq 0,\ell}
c(md,\ell) q^{am} y^{a\ell}.
\ee
Comparing with the expression (\ref{F}) for the free energy $\cal F$,
we thus observe that it can be rewritten as a sum of Hecke operators
acting on the elliptic genus of $M$
\be
\cF(p,q,y) = \sum_{N>0} \, p^N \; T_N 
\chi(M;q,y).
\ee
(See also \cite{borcherds,gritsenko} for similar expressions.)

This representation has a natural interpretation that arises 
from the geometric meaning of the Hecke operators $T_N$. 
The expression on the right-hand side of (\ref{deftn}) that
defined $T_N\phi$ can be reformulated as the sum of pullbacks 
for all holomorphic maps $f:T^2 \ra T^2$ of degree $N$
\be
\label{f}
T_N\phi = {1\over N} \sum_{f} f^*\phi.
\ee
These maps $f$ act as linear transformations on the two-torus and
can be represented by the matrices
\be
f = \twomatrixd a b 0 d.
\ee
where $ad=N$ and $0\leq b \leq d-1$. The factor $1/N$ in (\ref{f}) 
is natural because of the automorphisms of the torus.

With this interpretation, the free energy is represented as a sum over 
holomorphic maps 
\be
\cF(p,q,y) = \sum_{f:\,T^2 \ra T^2} {1\over N_f} p^{N_f} 
f^*\chi(M;q,y)
\ee
with $N_f$ the degree of the map $f$. The right-hand side can be
recognized as a summation over instanton sectors.

\newsubsection{Automorphic properties}

As suggested by its form, the above expression can indeed be
reproduced from a standard string one-loop computation.  To make this
correspondence precise, we notice that the partition function $\cZ$ is
in fact almost equal to an automorphic form for the group $SO(3,2,\Z)$
of the type discussed in \cite{borcherds}.
 
The precise form of this automorphic function has 
been worked out in detail in \cite{neumann}. It is defined by the product
\be
\Phi(p,q,y) = p^a q^b y^c \prod_{(n,m,\ell)>0}(1-p^n q^m 
y^\ell)^{c(nm,\ell)}
\ee
where the positivity condition means: $n,m\geq0$ with $\ell>0$ in the case 
$n=m=0$. The `Weyl vector' $(a,b,c)$ is defined by
\be
a=b={1\over 24} \chi(M),\qquad c=\sum_\ell -{|\ell|\over 4} c(0,\ell).
\label{weyl}
\ee
One can then show that the expression $\Phi$ is an automorphic form of
weight $c(0,0)/2$ for the group $O(3,2,\Z)$ for a suitable quadratic
form of signature $(3,2)$, see \cite{neumann}.

The form $\Phi$ follows naturally from a standard one-loop string
amplitude defined as an integral over the fundamental domain
\cite{harveymoore,kawai}.  The integrand consists of the genus one
partition function of the string on $M\times T^2$ and has a manifest
$O(3,2,\Z)$ T-duality invariance. We will not write down the explicit form of
this partition function, but refer to \cite{neumann} for the specific
details. For our purpose it is sufficient to mention the
final result of the integration 
\be
\cI = -\log \left(Y^{c(0,0)/2} |\Phi(p,q,y)|^2\right)
\ee
with $Y=\rho_2\sigma_2 - \half d \, \upsilon_2^2$, $d=\dim M$, in the notation
(\ref{exp}). Since the integral
$\cI$ is by construction invariant under the T-duality group $O(3,2,\Z)$,
this determines the automorphic properties of $\Phi$. The factor $Y$
transforms with weight $-1$, which fixes the weight of the form $\Phi$
to be $c(0,0)/2$.
 
The holomorphic contribution in $\cI$ is recovered by taking the limit
$\overline{p} \ra 0$. In the sigma model this corresponds to the
localization of the path-integral on holomorphic instantons and in
this way one makes contact with the description of the free energy
$\cF$ in the previous subsection. We note however that $\log\Phi$
contains extra terms that do not appear in $\cF$. Apart from a $\log
p$ contribution that arises from degree zero maps\footnote{ For degree
zero the two-torus gets mapped to a point in $M$, and the moduli space
of such maps is the product $M\times\cM_1$, with $\cM_1$ the moduli
space of elliptic curves. Weighting this contribution by the
appropriate characteristic class \cite{bcov}, we obtain
$-{\chi(M)\over 24} \log p,$ in accordance with (\ref{weyl}).}  these
terms are independent of $p$ and have no straightforward
interpretation in terms of instantons.

\newsection{Concluding Remarks}

Our computation of the elliptic genus of the symmetric product space
$\SNM$ can be seen as a refinement of the calculations in
\cite{hirzebruch,vafawitten} of the orbifold Euler number. In fact,
if we restrict to $y=1$, the elliptic genus reduces to the Euler number
and our identity takes the simple form
\be
\label{euler}
\sum_{N\geq 0} p^N \chi(\SNM) = \prod_{n>0} {1\over (1-p^n)}{}_{\chi(M)}.
\ee
Here the K\"ahler condition is not necessary. If $M$ is an algebraic
surface, it can be shown that this formula also computes the {\it
topological} Euler characteristic of the Hilbert scheme $M^{[N]}$ of
dimension zero subschemes of length $n$ \cite{goettsche}.  This space
is a smooth resolution of the symmetric product $\SNM$.  (In complex
dimension greater than two the Hilbert scheme is unfortunately not
smooth.) It is natural to conjecture that in the case of a
two-dimensional Calabi-Yau space, \ie a $K3$ or an abelian
surface, the orbifold elliptic genus of the symmetric product 
also coincides with elliptic genus of the Hilbert scheme.

The left-hand side of our identity (\ref{identity}) can be seen to compute
the superdimension of the infinite, graded vector space
\be
\bigoplus_{N,m,\ell} V_{m,\ell}(\SNM),
\ee
where $V_{m,\ell}$ are the index bundles (\ref{index}). Our result
suggests that this space forms a natural representation of the
oscillator algebra generated by string field theory creation operators
$\alpha_{n,m,\ell}^I$.  This statement is analogous to the assertion
of Nakajima \cite{nakajima} (see also \cite{groj}) that the space
$\oplus_N H^*(M^{[N]})$ forms a representation of the Heisenberg
algebra generated by $\alpha_n^I$, where $I$ runs over a basis of
$H^*(M)$.

It would be interesting to explore possible applications to gauge
theories along the lines of \cite{vafawitten}. On a $K3$ manifold the
moduli space of Yang-Mills instantons takes (for certain instanton
numbers) the form of a symmetric product of $K3$. This fact was used
in \cite{vafawitten} to relate the partition function of $N=4$
Yang-Mills theory on $K3$ to the generating function of Euler numbers
(\ref{euler}).  Our formula gives an explicit expression for the
elliptic genus of these instanton moduli spaces.  It seems a natural
conjecture that the analysis of \cite{vafawitten} can be generalized
to show that the generating function of the elliptic genera is the
partition function of an appropriately twisted version of $N=2$
Yang-Mills theory on $K3\times T^2$. For some interesting
recent work in this direction, see \cite{nekrasov}.

Finally, our calculation is likely to be relevant for understanding
the quantum statistical properties of D-branes \cite{polch} and their
bound states \cite{witten}.  Particularly useful examples of such
possible bound states are those between D-strings with one (or more)
higher dimensional D-branes.  In type II string compactifications on
manifolds of the form $\MM\times S^1$, we can consider the
configuration of a D-string wound $N$ times around the $S^1$ bound to
a (dim$M$+1)-brane.  (For the case where $M$ is a $K3$ manifold, this
situation was first considered by Vafa and Strominger
\cite{stromingervafa} in their D-brane computation of the
5-dimensional black hole entropy.)  As argued in
\cite{vafa6d,stromingervafa}, the quantum mechanical degrees of
freedom of this D-brane configuration are naturally encoded in terms
of a two-dimensional sigma model on the $N$-fold symmetric tensor
product of $M$, that describes the transversal fluctuations of the
D-string. As was also pointed out in \cite{maldacenasusskind}, this
description implies that a multiply wound D-string can carry
fractional oscillation numbers. Our result shows that the
resulting quantum statistical description of these first quantized
``fractional'' D-strings is in fact equivalent to a description in
terms of second quantized ``ordinary'' strings. In this correspondence
the extra degrees of freedom that arise from the fractional moding 
are used to assign to each individual string a momentum along
the $S^1$ direction. This result may be a useful clue in explaining 
some of the miraculous non-perturbative dualities between strings and 
D-branes.

\vspace{15mm}

{\noindent \sc Acknowledgements}

We thank D. Neumann, J-S. Park, G. Segal, W. Taylor and C. Vafa for
discussions, and the Aspen Center of Physics for hospitality during
the final stage of this work. This research is partly supported by a
Pionier Fellowship of NWO, a Fellowship of the Royal Dutch Academy of
Sciences (K.N.A.W.), the Packard Foundation and the A.P. Sloan
Foundation.

\newpage

%\vspace{15mm}
%\pagebreak[3]

\renewcommand{\thesection}{A}
\renewcommand{\thesubsection}{A.\arabic{subsection}}
\addtocounter{section}{1}
\setcounter{equation}{0}
\setcounter{subsection}{0}
\setcounter{footnote}{0}
\begin{center}
{\sc Appendix: Elliptic Genus}
\end{center}

\medskip
 
We summarize some facts about the elliptic genus for a K\"ahler
manifold $M$ of complex dimension $d$ \cite{book,ell}.  We start with
an elliptic curve $E$ with modulus $\tau$ and a line bundle labeled by
$z\in Jac(E)\cong E$. We define $q=e^{2\pi i \tau}, y = e^{2\pi i
z}$. The elliptic genus is defined as
\be
\chi(M;q,y)=\Tr_{\strut \cH(M)}(-1)^Fy^{F_L} q^{L_0-{d\over 8}} 
\qbar^{\Lbar_0-{d\over 8}}, 
\ee
where $F=F_L+F_R$ and  $\cH(M)$ is the Hilbert space of the $N=2$ 
supersymmetric field theory with target space $M$. 

For a Calabi-Yau space the elliptic genus is a 
weak Jacobi form of weight zero and index $d/2$.  Recall that a Jacobi form
$\phi(\tau,z)$ of weight $k$ and index $r$ (possibly half-integer) 
transforms as \cite{eichler}
\ba
\phi\left({a\tau + b \over c\tau +d},{z\over c\tau+d}\right) \is
(c\tau + d)^k e^{\pi i {rcz^2\over c\tau + d}} \phi(\tau,z),\nonu
\phi(\tau,z + m\tau + n) \is e^{-\pi i r(m^2\tau + 2 m z)}\phi(\tau,z),
\ea
and is called weak if it has a Fourier expansion of the form
\be
\phi(\tau,z) = \sum_{m\geq 0,\ell} c(m,\ell) q^my^\ell.
\ee
The coefficients of such a form depend only on $4rm - \ell^2$ and on
$\ell$ mod $2r$.

The elliptic genus has the following properties: First of all, it is
a genus; that is, it satisfies the relations
\ba
\chi(M \sqcup M';q,y)\is \chi(M;q,y)+ \chi(M';q,y), \nonu 
\chi(M\times M';q,y) \is \chi(M;q,y)\cdot \chi(M';q,y)\\[2mm]
\chi(M;q,y) \is 0,\qquad\mbox{if $M=\partial N$,} \nonumber
\ea
where the last relation is in the sense of complex bordism.
Furthermore, for $q=0$ it reduces to a weighted sum over the Hodge numbers,
which is essentially the Hirzebruch $\chi_y$-genus,
\be
\chi(M;0,y)=\sum_{p,q} (-1)^{p+q}y^{p-{d\over 2}} h^{p,q}(M),
\ee
and for $y=1$ its equals the Euler number of $M$
\be
\chi(M;q,1)=\chi(M).
\ee

For smooth manifolds, the elliptic genus has an alternative definition
in terms of characteristic classes, as follows.  For any vector bundle
$V$ one defines the formal sums
\be
\ext_q V=\bigoplus_{k\geq0}  q^k \ext^k V, \qquad
S_qV= \bigoplus_{k\geq0} q^k S^k V, 
\ee
where $\ext^k$ and $S^k$ denote the $k$th exterior and symmetric product 
respectively.  One then has an equivalent definition of the elliptic genus 
as
\be
\chi(M;q,y)=\int_M  ch(E_{q,y}) td(M)
\ee
with
\be
E_{q,y} =  y^{-{d\over 2}} \bigotimes_{n\geq 1}\Bigl( \ext_{-yq^{n-
1}}T_M\otimes \ext_{-y^{-1}q^n} \overline{T}_M \otimes S_{q^n} T_M\otimes 
S_{q^n} \overline{T}_M\Bigr)
\ee
where $T_M$ denotes the holomorphic tangent bundle of $M$. Expanding the
bundle $E_{q,y}$ as
\be
E_{q,y} = \bigoplus_{m,\ell} q^m y^\ell E_{m,\ell}
\ee
one can define the coefficients $c(m,\ell)$ as
\be
c(m,\ell) = \hbox{index} \Dslash_{E_{m,\ell}}
\ee
with $\Dslash_E$ the Dirac operator twisted with the vector bundle $E$.
So $c(m,\ell)$ computes the dimension of the virtual vector space
\be
V_{m,\ell}(M) = {\rm ker} \Dslash_{E_{m,\ell}} \ominus
{\rm cok} \Dslash_{E_{m,\ell}}
\label{index}
\ee

\renewcommand{\Large}{\large}

\end{document}